\documentclass[showpacs,amsmath,amssymb,aps,10pt,reprint,superscriptaddress,prb]{revtex4-1}
\usepackage{graphicx}
\usepackage{bm}
\usepackage[breaklinks=true,colorlinks=true,linkcolor=blue,urlcolor=blue,citecolor=blue]{hyperref}
\usepackage{graphics}
\usepackage{dcolumn}
\usepackage{amsmath,amssymb}
\usepackage{mathdots}
\usepackage{natbib}
\usepackage{soul,color}
\usepackage{epstopdf}
\usepackage[note-name]{notes2bib}
\begin{document}

\title{Effect of electron irradiation and Pr doping on the charge transport in YBCO single crystals}

\author{R. V. Vovk}
\author{G. Ya. Khadzhai}
\author{O.~V.~Dobrovolskiy}
    %\email[Corresponding author: ]{Dobrovolskiy@Physik.uni-frankfurt.de}
    \affiliation{Physikalisches Institut, Goethe University, 60438 Frankfurt am Main, Germany}
    \affiliation{Physics Department, V. Karazin Kharkiv National University, 61077 Kharkiv, Ukraine}

\begin{abstract}
The influence of irradiation by electrons with energies of $0.5 ... 2.5$\,MeV at temperatures of about $10$\,K on the basal-plane resistivity of the YBa$_2$Cu$_3$O$_{7-\delta}$ single crystals is investigated in the range from $T_c$ to $300$\,K. The resistivity temperature dependence is determined by defects arising due to the irradiation. These defects directly affect the superconducting transition, decreasing $T_c$ and increasing the transition width without significant distortions of its shape. The resulting defects also lead to an increase in the Debye temperature due to a reduction of the anisotropy, and a noticeable increase in the scattering by phonons in the sample. The excess conductivity does not change with the irradiation used.
\end{abstract}

\maketitle
%% \linenumbers

\enlargethispage{2\baselineskip}
\section{Introduction}
Synthesis of new superconducting materials with tailored current-carrying properties via modification of their structure and charge transport characteristics \cite{Dob17nsr,Sol16prb,Dob15met} is an actual research line in modern condensed matter physics. One of the most efficient experimental approaches allowing one to accomplish this task is irradiation and processing of the investigated samples by electron beams \cite{Gia94prb,Kha18jms,Hut18mee,Dob15bjn} as well as modification of their elemental composition \cite{Bor91ssc,Vov07jms}. In this respect, cuprates from the high-$T_c$ superconducting 1-2-3 system RBa$_2$Cu$_3$O$_{7-\delta}$ (R = Y or other rare earths) are most asked-for for several reasons. Firstly, these compounds have a rather high critical temperature $T_c$ \cite{Wum87prl}, above the nitrogen liquefaction temperature. Secondly, there are well-developed fabrication techniques for the fabrication of polycrystalline \cite{Bon17ltp} and cast solid-phase samples \cite{Lot10ltp} of this compound of rather large sizes. Finally, the physics characteristics of this compound can relatively easily be tuned by complete \cite{Kla98pcs,Vov15pcs} or a partial \cite{Vov15jms,Vov09phb} isovalent or non-isovalent substitution. At the same time, the presence of a labile component in the compound can cause a nonequilibrium state in samples of a non-stoichiometry composition with respect to oxygen \cite{Met93pcs,Vov11jms}. Such a state can easily be induced by application of a high pressure \cite{Sad00prb,Bal97ltp}, an abrupt temperature jump \cite{Jor90pcs} or appear in consequence of a long-term storage or aging \cite{Mar95apl,Vov14jms1}, In this respect, the use of electron irradiation \cite{Rul03prl} is rather versatile. For instance, an irradiation of YBa$_2$Cu$_3$O$_{7-\delta}$ with fast electrons leads to a linear increase of the electrical resistance and a linear decrease of the transition temperature without significant volume deviations from the oxygen stoichiometry \cite{Pet91phb}.

Among the approaches to change the elemental composition of this compound, one should note a complete \cite{Akh02phb,Vov14ssc} or a partial \cite{Chr07acs} substitution of Y by Pr. The latter, in contrast to other rare earths \cite{Mur14pcs}, allows for the $T_c$ variation from $0$ to the maximum value while keeping the oxygen doping degree at the optimal level \cite{Akh02phb,Vov14ssc,Chr07acs}. I this way, the use of electron irradiation and Pr doping allows for tuning the critical and electrical transport properties of RBa$_2$Cu$_3$O$_{7-\delta}$ single crystals in a broad range. Simultaneously, the stability of the oxygen subsystem is remaining intact, that is especially important in view of technological applications of the compound \cite{Bon17ltp}.
\enlargethispage{2\baselineskip}

At the same time, these techniques may noticeably affect the characteristics of the temperature dependences of the electrical conductivity, thus determining the conditions for the realization of a series of non-trivial phenomena peculiar to high-$T_c$ compounds in the normal, non-superconducting state. These phenomena include the so-called pseudogap anomaly \cite{Sad05prb}, fluctuation paraconductivity \cite{Gho11mpl}, metal-insulator transition \cite{Gin89boo,Vov13pcs}, incoherent electric charge transport \cite{And91prl} and so on, that possess a rich physics from the viewpoint of the academic research. Indeed, according to contemporary views \cite{Sol16prb,Akh02phb} it is these phenomena which may shed light on our understanding of the microscopic nature of high-$T_c$ superconductivity, which remain unresolved despite more than three decades of extensive experimental and theoretical investigations.

In a previous work we investigated the effect of electron irradiation on the basal-plane conductivity of an optimally doped YBCO single crystal with $T_c = 91.74$\,K \cite{Aza17jms}. A noticeable increase of its electrical resistance in consequence of the electron irradiation has been observed in the broad temperature range $T_c - 300$\,K. Here, we compare the effects of electron irradiation and praseodymium doping on the charge scattering parameters and the superconducting properties of the same single crystal.

\section{Experimental}
The samples were grown by the solution-melt technique. After the growth the samples were saturated with oxygen at 430$^\circ$C during 4 days. All the investigated samples were twinned, while the twin planes had a block structure. The typical sample dimensions were $1.5..2\times0.2..0.3\times0.01..0.02$\,mm$^3$, where the smallest dimension corresponds to the $c$ axis. The transport current was applied along the longest side of the sample, while the electrical resistance was measured in the standard four-probe geometry, with a distance between the voltage contacts of $1$\,mm.

Irradiation was done with electron energies $0.5...2.5$\,MeV at $T \lesssim 10$\,K. The sample temperature during the irradiation process did not exceed $10$\,K. The irradiation dose was $8.8 \times10^{18}$\,cm$^{-2}$ corresponding to a defect concentration of $10^{-4}$\,dpa \cite{Bon01prb}. A specially designed He cryostat allowed for resistance measurements directly after the electron irradiation in the temperature range $10$\,K$<T<500$\,K. All measurements were done at a fixed temperature. The temperature was measured with a platinum resistance thermometer. The temperature stability was better than $5\,$mK.

\section{Results and discussion}
\subsection{Normal resistance}
The temperature dependence of the basal-plane resistivity, $\rho_{ab}(T)$, was described taking into account scattering of the charge carriers on phonons and defects, as well as the excess conductivity
\begin{equation}
    \label{e1}
    \rho_{nab} = \frac{1}{\frac{1}{\rho_0 + \rho_{ph}} + b_1 (e^{T_1/T} - 1)}.
\end{equation}
\enlargethispage{2\baselineskip}

\begin{figure}[t!]
    \centering
    \includegraphics[width=0.85\linewidth]{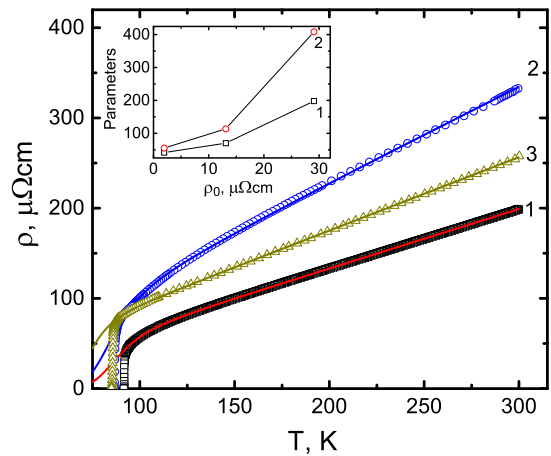}
    \caption{Temperature dependences of the resistance of YBa$_2$Cu$_3$O$_{7-\delta}$ and Y$_{1-y}$Pr$_y$Ba$_2$Cu$_3$O$_{7-\delta}$ single crystals. 1 --- initial state, 2 --- after electron irradiation, 3 --- Y$_{1-y}$Pr$_y$Ba$_2$Cu$_3$O$_{7-\delta}$. Lines --- fits to Eqs. (\ref{e1}) and (\ref{e2}). Inset: Dependence of the fitting parameters to Eqs. (\ref{e1}) and (\ref{e2}) on the residual resistivity: 1 --- $\theta$,\,K, 2 --- $C_3$,\,$\mu\Omega$cm.}
    \label{f1}
\end{figure}

Here $\rho_0$ characterizes charge carriers scattering on defects and
\begin{equation}
    \label{e2}
    \rho_{ph} = C_3\left(\frac{T}{\theta}\right)^3\int_0^{\theta/T} \frac{e^x x^3 dx}{(e^x - 1)^2}
\end{equation}
is the Bloch-Gr\"uneisen expression \cite{Col65jap}. The term $b_1 (e^{T_1/T})$ describes the excess conductivity \cite{Aza17jms}. The respective fits are shown in Fig. \ref{f1} by solid lines. The fitting parameters providing a fitting error within $1\%$ are reported in Table \ref{t1}.

From Table \ref{t1} it follows that electron irradiation causes a factor of 15 increase of the residual resistivity, $\rho_0$, that is it leads to a strong enhancement of the disorder degree in the sample. It is important to emphasize that a similar suppression of $T_c$ for electron irradiation is accompanied by a factor of two larger increase of $\rho_0$ than for Pr doping. This means that electron irradiation generates an essentially broader spectrum of defects than the introduction of Pr. At the same time, only a part of defects caused by the irradiation has effect on $T_c$.

From Table \ref{t1} and the inset of Fig. \ref{f1} it follows that $T_c$ non-monotonically depends on the residual resistivity, whereas the fitting parameters to Eqs. (\ref{e1}) and (\ref{e2}) increase with increasing $\rho_0$ monotonically.

The small value of the Debye temperature, $\theta$, in the initial state of the sample is likely caused by the anisotropy of the sample. We attribute this to that the interaction between the layers is much weaker than within the layers so that $\theta$ associated with the transverse vibrations along the $c$-axis is much smaller than $\theta$ related to the transverse vibrations within the layers \cite{Ans88etp}. The increasing degree of disorder leads to an isotropization of the phonon spectrum that, in turn, causes an increase of the Debye temperature. We note that the Debye temperature averaged over the elements taking into account the stoichiometry only amounts to $\langle\theta \rangle\approx 345$\,K. The phonon resistance coefficient, $C_3$, also increases after irradiation, that agrees well with the data for transient metals \cite{Kho83fnt}.

The electron irradiation has almost no effect on the parameters $b_1$ and $T_1$ characterizing the excess conductivity. Therefore we assume that the excess conductivity only weakly depends on the degree of disorder in the sample.

\subsection{Superconducting transition}
\begin{figure}[t!]
    \centering
    \includegraphics[width=0.9\linewidth]{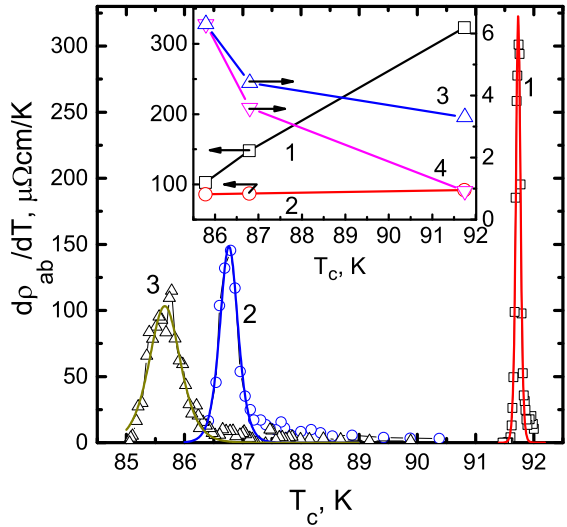}
    \caption{Derivatives $d\rho/dT$ in the region of the superconducting transition. The curve numbering corresponds Fig. \ref{f1}. The lines correspond to Eq. (\ref{e3}). Inset: Dependences of the parameters for Eqs. (\ref{e3}) and (\ref{e4}) on $T_c$: 1 --- $d\rho/dT|_{T= T_c}$,\,$\mu\Omega$cm/K, 2 --- $T_{cross}$,\,K, 3 --- $10^3\varepsilon_{cross}$, 4 --- $10 \Delta T_{c0.5}$,\,K.}
    \label{f2}
\end{figure}

The superconducting transition temperature was determined at the low-temperature maximum of the derivative $d\rho/dT$. Figure \ref{f2} illustrates the curves $d\rho/dT$ in the temperature range of the superconducting transition. The derivatives were fitted to the expression \cite{Rol83boo}
\begin{equation}
    \label{e3}
    \frac{d\rho(T)}{dT} = \frac{\rho_1 e^{-Z}}{w(1+e^{-Z})^2},
\end{equation}
where $w$ and $\rho_1$ are parameters.

\begin{table*}[t!]
\centering
\begin{tabular}{|c|c|c|c|}
\hline
                & YBa$_2$Cu$_3$O$_{7-\delta}$             & YBa$_2$Cu$_3$O$_{7-\delta}$         & Y$_{1-x}$Pr$_x$Ba$_2$Cu$_3$O$_{7-\delta}$\\
                & before  & after   & $x = 0.05$\\
                & irradiation & irradiation & \\
                \hline
$T_c$, K           & 91.74             & 83.79               & 85.78\\
$\rho_0$, m$\Omega$cm & 1.95          & 29.05            & 13.08\\
$\theta$, K           & 41.5             & 198              & 70\\
$C_3$, m$\Omega$cm    & 54.62           & 408.43            & 113.75\\
$T_1$, K              & 1132            & 1065              & -\\
$b_0$, (m$\Omega$cm)$^{-1}$  & $3.20\times10^{-8}$ &$3.20\times10^{-8}$&  -\\
$\Delta T_{c0.5}$, K & 0.092          & 0.36               & 0.63\\
$d\rho/dT|_{T = T_c}$, $\mu\Omega$cm/K & 323          & 148               & 103\\
$\varepsilon_{cross}$ & 0.0033          & 0.0044               & 0.0063\\
$T_{cross}$, K        & 92.04     & 87.26                & 86.32\\
\hline
\end{tabular}
   \caption{Fitting parameters for $\rho_{nab}(T)$ to Eqs. (\ref{e1}) and (\ref{e2}).}
   \label{t1}
\end{table*}

From Figure \ref{f2} and Table \ref{t1} it follows that the width of the superconducting transition at the half height, $\Delta_{c0.5}\approx 3.5w$, has noticeably increased, the maxima of the derivative heights, $d\rho/dT|_{T = T_c} = \rho_1/(4w)$ have considerably decreased, and the maxima have remained symmetric. Such modifications attest to that in consequence of electron irradiation a certain amount of defects has appeared, that affect the superconducting transition, but their spatial distribution is macroscopically homogeneous.

\subsection{Fluctuation conductivity}
Near the superconducting transition, the fluctuation conductivity in the plane of the layers is described by the expression \cite{Lar09boo}
\begin{equation}
    \label{e4}
    \sigma_{ab} = \frac{e^2}{16\hbar d}\frac{1}{\sqrt{\varepsilon(\varepsilon + T)}},
\end{equation}
where $d = 11.7\,\mathrm{\AA}$ is the interlayer distance \cite{Keb89prb}, $\varepsilon = \frac{T - T_c}{T_c}\ll1$, and $r =(4(_\downarrow c^\uparrow(0)))/d^\uparrow 2$. Equation (\ref{e4}) describes the 2D-3D crossover occurring in some temperature range: At $\varepsilon\ll r$ one has the 3D regime with $\sigma_{ab} \propto (\varepsilon r)^{-1/2}$, whereas at $\varepsilon\gg r$ the 2D regime is realized with $\sigma_{ab} \propto \varepsilon^{-1}$. The crossover point $r = \varepsilon$ separates the 2D and 3D regimes.

Near $T_c$, where $\varepsilon \ll 1$ and $\sigma_{ab}(\varepsilon) \propto \varepsilon^{-1/2}$, one can assume that $\sigma_{ab}(T)\ll 1/\rho_{ab}(T)$ and one can neglect the normal conductivity associated with the scattering on impurities and phonons. Then $\sigma_{ab}(T) \approx 1 /\rho_{ab}(T)$ and from Eq. (\ref{e4}) one can deduce the anisotropy parameter, $r$, for different $\varepsilon \ll 1$ at the investigated experimental conditions. For these calculations we used the experimental values $\rho(T)$ which belong to the temperature intervals corresponding to the right-hand slopes of the $d\rho(T)/dT$ maxima.

The inset of Fig. \ref{f2} illustrates the dependences of the characteristics of the superconducting transition according to Eqs. (\ref{e3}) and (\ref{e4}) on $T_c$. One sees that these dependences are monotonic. In this way, the characteristics of the normal resistance monotonically depend on the residual resistance (that is on the total number of defects), while the characteristics of the superconducting transition monotonically depend on the superconducting transition temperature which is determined by the concentration of oxygen vacancies in the Cu-O(2) layers.

\begin{figure}[t!]
    \centering
    \includegraphics[width=0.87\linewidth]{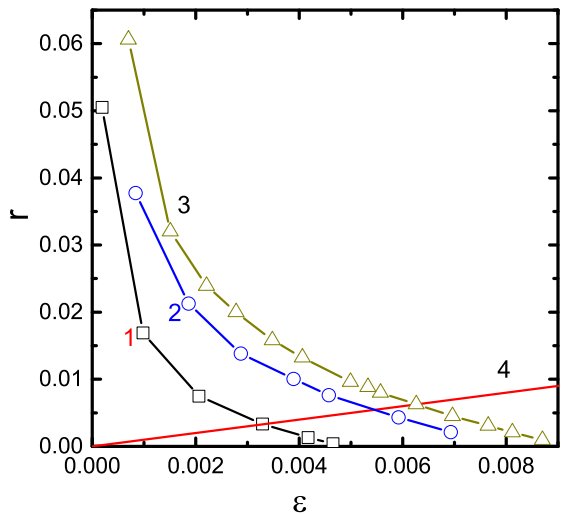}
    \caption{Dependence of the anisotropy parameter $r$ on the reduced temperature $\varepsilon$. 1 --- initial state, 2 --- after irradiation, 3 --- Y$_{1-y}$Pr$_y$Ba$_2$Cu$_3$O$_{7-\delta}$, 4 --- crossover line $r = \varepsilon$.}
    \label{f3}
\end{figure}

The values of $r$ are reported in Fig. \ref{f3} along with the crossover line $r =\varepsilon$. One sees that at $\varepsilon \leq 0.01$ the parameter $r$ monotonically decreases with increasing $\varepsilon$, that is at $\varepsilon \leq 0.01$ there is a dependence $r(\varepsilon)$ for the studied experimental conditions. Even for $T \rightarrow T_c$ these dependences $r(\varepsilon)$ have no intersections, but rather they are shifted up with respect to each other in accordance with the reduction of the superconducting transition temperature. The crossover line $r = \varepsilon$ crosses all curves $r(\varepsilon)$. The values of $\varepsilon_{cross}$ increase with decreasing $T_c$, as shown by curve 3 in the inset of Fig. \ref{f2}.

From Fig. \ref{f2}, the insets in Fig. \ref{f3} and Table \ref{t1} it follows that the crossover temperatures determined in this way are far enough from $T_c$ as $T_{cross} > (T_c + 0.5 \Delta T_{c0.5})$. Thus, in the temperature range corresponding to the right-hand slope of the $d\rho/dT$ maxima the motion of charge carriers is three-dimensional. This regime corresponds to the region $\varepsilon < r$ lying above the crossover line $\varepsilon =r$ in Fig. \ref{f3}. It is worth noting that an increase of $r$ upon a decrease of $T_c$ means an increase of $\xi_c(0)$ owing to the increase of the number of oxygen vacancies. However, the usage of the relation $r =(4(_\downarrow c^\uparrow(0)))/d^\uparrow 2$ for the determination of the coherence length yields $\xi_c(0)\approx 0.5\,\mathrm{\AA} \ll d$ that agrees well with the literature data \cite{Fri89prb} but is not consistent with the 3D regime, where $\xi_c(0) >d$. This contradiction may be stipulate by the presence of a non-superconducting phase causing an inhomogeneous current distribution in the sample \cite{Ohb88prb,Sol09fnt}. We note that the presence of the non-superconducting phase affects strongly the $\rho_{exp}(T)$ value than the shape and the width of the $d\rho/dT$ maxima. A certain role at this can be played by other specific mechanisms of quasiparticle scattering \cite{Apa02prb65,Vov03prb,Ada94ltp,Vov03prl,Cur11prb} stipulated by the structural and kinematic anisotropy in the system.

Summing up, the results of our study suggest that while electron irradiation leads to the appearance of a considerable number of defects in the sample, only some part of them has an affect on the superconducting transition, increasing its width and reducing $T_c$. The presence of defects noticeably reduces the anisotropy of the sample and enhances charge carriers scattering on phonons. At the same time, the excess conductivity is not affected by the employed electron irradiation.

\section*{Acknowledgements}
The research leading to these results has received funding from the European Union's Horizon 2020 research and innovation program under Marie Sklodowska-Curie Grant Agreement No. 644348 (MagIC).

%\bibliographystyle{unsrt}

% \bibliographystyle{elsarticle-num}
% \bibliographystyle{elsarticle-harv}
% \bibliographystyle{elsarticle-num-names}
% \bibliographystyle{model1a-num-names}
% \bibliographystyle{model1b-num-names}
% \bibliographystyle{model1c-num-names}
% \bibliographystyle{model1-num-names}
% \bibliographystyle{model2-names}
% \bibliographystyle{model3a-num-names}
% \bibliographystyle{model3-num-names}
% \bibliographystyle{model4-names}
% \bibliographystyle{model5-names}
% \bibliographystyle{model6-num-names}

%\bibliography{D:/bibliobase/ybco,D:/bibliobase/fluxonics,D:/bibliobase/magnonics,D:/bibliobase/proximity}

\end{document}